\documentclass{article}
\usepackage{graphicx}

\evensidemargin=\oddsidemargin
\def\Title#1#2#3{%
    \baselineskip=18pt
    \begin{center}
          {\large\bf{#1} \\ }
          \bigskip\bigskip
          {#2} \\
          {#3} \\
    \end{center}}
\long\def\Abstract#1{%
         \bigskip
         \parbox{0.93\textwidth}{%
                 \begin{center}
                       {\bf Abstract} \\
                 \end{center}
                 \medskip{\baselineskip=14pt #1}
                 \vss}
         \bigskip}

\makeatletter
\renewcommand{\section}%
 {\@startsection{section}{1}{0pt}%
  {-3.25ex plus -1ex minus -.2ex}{1.5ex plus .2ex}%
  {\vspace*{5mm}\raggedright\large\bf }}
\renewcommand{\subsection}%
 {\@startsection{subsection}{2}{0pt}%
  {-2.25ex plus -.5ex minus -.2ex}{-1.5ex plus -.2ex}{\bf }}
\renewcommand{\subsubsection}%
 {\@startsection{subsubsection}{3}{0pt}%
  {-1.25ex plus -.2ex minus -.1ex}{-1.2ex plus -.2ex}{\bf }}

\makeatother

\begin{document}

\Title{On the meaning of the wave function of the Universe}%
{T. P. Shestakova}%
{Department of Theoretical and Computational Physics,
Southern Federal University,\\
Sorge St. 5, Rostov-on-Don 344090, Russia \\
E-mail: {\tt shestakova@sfedu.ru}}

\Abstract{The meaning of the wave function of the Universe was actively discussed in 1980s. In most works on quantum cosmology it is accepted that the wave function is a probability amplitude for the Universe to have some space geometry, or to be found in some point of the Wheeler superspace. It seems that the wave function gives maximally objective description compatible with quantum theory. However, the probability distribution does not depend on time and does not take into account the existing of our macroscopic evolving Universe. What we wish to know is how quantum processes in the Early Universe determined the state of the present Universe in which we are able to observe macroscopic consequences of these quantum processes. As an alternative to the Wheeler -- DeWitt quantum geometrodynamics we consider the picture that can be obtained in the extended phase space approach to quantization of gravity. The wave function in this approach describes different states of the Universe which correspond to different stages of its evolution.}

\section{Introduction}
In 1980s there was a peak of interest to quantum cosmology which partially continued in 1990s (see, for example, \cite{HH,Vil1,Vil2,HT,Linde}). Most papers were based on the Wheeler -- DeWitt quantum geometrodynamics \cite{DeWitt} the main object of which is a wave function of the Universe that can be defined as a probability amplitude for the Universe to have some space geometry and some matter distribution. It was declared \cite{Hartle} that the wave function describes an initial state for the subsequent classical evolution of the Universe. In a widely accepted cosmological scenario a closed isotropic universe is considered which spontaneously nucleates from a classically forbidden region of minisuperspace at some value of the scale factor. It seems that the wave function gives the probability amplitude at the moment of nucleation.

On the other hand, the wave function satisfies the Wheeler -- DeWitt equation and, therefore, does not depend on time. On this occasion DeWitt wrote \cite{DeWitt}:
\begin{quote}
``\ldots we shall interpret it as informing us that the coordinate labels $x^{\mu}$ are really irrelevant. Physical significance can be ascribed only to the intrinsic dynamics of the world\ldots''.
\end{quote}
Then, if we accept that quantum theory is universal and applicable to any object including our macroscopic Universe, the wave function is not the probability amplitude at the moment of nucleation but rather a probability amplitude that is true at every moment of the Universe existence.

At this point we meet if not a contradiction then at least an ambiguity. If the wave function not depending on time predicts that small values of the scale factor are more probable, the Universe describing by this wave function has no chance to become macroscopic. Otherwise, if the wave function predicts that large values of the scale factor are more probable, it means just the statement of the fact that our Universe is macroscopic, but it would not tell us anything about the early stage of its existence. Then, since space geometry changes (values of the scale factor are different at different moments of time) in our macroscopic Universe, it would be reasonably to assume that the probability distribution must also change with time, but it contradicts to the basic statements of the Wheeler -- DeWitt quantum geometrodynamics.

As far as I know, this question has not been discussed in the literature on quantum cosmology. Most authors tacitly accept that the wave function describes the Early Universe, but make use of the Wheeler -- DeWitt equation to find it. Also, the next question can be posed: do we seek for a wave function which would predict that our Universe, as we observe it now, is most probable, so that quantum cosmology justifies, in a certain sense, the existence of our Universe? Or, do we seek some ``objective'' probability distribution which would help us to understand how probable a universe like ours is? The second possibility can hardly be realized. The reason is that there is a variety of solutions of the Wheeler -- DeWitt equation and one needs additional arguments to single out the only suitable solution or a class of such solutions. The solution can be singled out by boundary conditions which, as believed by some authors, must have the status of a fundamental physical law \cite{Hartle}.

The most-known proposals are the ``no boundary'' and tunneling boundary conditions \cite{HH,Vil1}. These boundary conditions are grounded on different rationales, but the authors of the both proposals claimed that their boundary conditions predict an inflationary stage of the Universe evolution. In other words, the true goal of quantum cosmology is to give a good reason for the inflationary stage which, in its turn, helps to solve some problems of classical cosmology. And again, in this case one would hardly seek for a ``global'' wave function describing the Universe at any time but rather wish to obtain a wave function which would provide desirable conditions in the beginning of the Universe.

The contradiction between the independence of the wave function on time and our aspiration to understand what phenomena in the Very Early Universe gave rise to the Universe which we observe now can be resolved in the framework of the extended phase space approach to quantization of gravity \cite{SSV1,SSV2,SSV3,SSV4} which was presented, in particular, at the previous conference ICPPA-2017 \cite{Shest1}. In the extended phase space approach the wave function satisfies the temporal Schr\"odinger equation instead of the Wheeler -- DeWitt equation and evolves with time, so that the Universe may have been in some state at the moment of its nucleation, in another state during inflation, and can be described by a quasiclassical wave function latter.

In Section 2 I shall remind what grounds we have to consider the Schr\"odinger equation as more fundamental than the Wheeler -- DeWitt equation in quantum gravity and what are the main features of the extended phase space approach (this question is discussed in details in \cite{Shest1}). In Section 3 an outline of a cosmological scenario will be given that can be proposed in the framework of this approach. Since in this scenario the choice of gauge conditions is important, in Section 4 the role of gauge conditions and their interpretation is considered. Section 5 comprises the main conclusions.

\section{What equation must the wave function of the Universe satisfy?}
Strictly speaking, until we do not have a consistent quantum theory of gravity which can be verified by observational data, we cannot be sure what equation the wave function of the Universe must satisfy. One can just give mathematical, logical or philosophical arguments in favor of some equations. The general opinion inclines to the belief that the wave function must satisfy the Wheeler -- DeWitt equation. However, there are not many arguments for it. As well-known, DeWitt \cite{DeWitt} grounded it on the Dirac quantization scheme \cite{Dirac1,Dirac2,Dirac3} which was the first attempt to construct a quantum theory of constrained systems. In this approach all variables are divided into physical and non-physical ones, only the former ones being included into (physical) phase space. It is natural to say that physical variables describe a ``physical'' subsystem while the role of gauge (non-physical) degrees of freedom is questionable. Dirac postulated that after quantization any constraint must become a condition on a state vector. But this postulate has been never confirmed by direct experiments, since very successful and experimentally verified gauge theories like quantum electrodynamics were based on other theoretical methods than the Dirac scheme. It seems to be paradoxical that the Dirac quantization scheme played no role in the creation of these successful theories and, at the same time, the result of its application to gravity, the Wheeler -- DeWitt equation, is recognizes as a fundamental one.

After Dirac other approaches to quantization of gauge fields have been developed which are grounded on the Feynman formulation \cite{Feyn} of quantum theory. Since, for a consistent theory, one can expect that different approaches would lead to the same results, attempts were made to derive the Wheeler -- DeWitt equation from a path integral. Various forms of the path integral were used as a starting point for this procedure. So, Barvinsky and Ponomariov \cite{BP} used the path integral over the so-called reduced phase space while Halliwell \cite{Hall} relied upon the formalism of extended phase space.

The idea of extended phase space was put forward by Batalin, Fradkin and Vilkovisky (BFV) \cite{BFV1,BFV2,BFV3} whose approach to quantization of gauge fields was recognized as most powerful since it can be applied to any gauge fields including those with open algebras. For our purposes it is of importance because extended phase space includes gauge and ghost (non-physical) degrees of freedom and we intend to analyse the role of gauge variables in quantum gravity. However, in the BFV approach gauge variables were still considered as non-physical, secondary degrees of freedom playing just an auxiliary role in the theory. The Hamiltonian form of action was constructed in such a way that the Hamiltonian coincides with the one built by the prescription of Dirac.

However, there exist another way to construct Hamiltonian dynamics of a constrained system exploiting the idea of extended phase space. The main source of difficulties with the Hamiltonian formulation was the impossibility to construct the Hamiltonian according to the usual rule
$H=p\dot q-L$, since for gauge variables their generalized velocities are missing in the Lagrangian and could not be expressed in coordinates and momenta. But the notion of extended phase space came from the path integral formulation of modern field theory, where the gauge-invariant action of the original theory is replaced by an effective action that includes gauge-fixing and ghost terms. As we shall see below, making use of a differential form of gauge conditions enables us to introduce the missing velocities into the Lagrangian. At this point our approach differs from the BFV one.

A special feature of the mentioned papers \cite{BP,Hall} is the consideration of the path integrals under asymptotic boundary conditions for ghosts and Lagrange multipliers of gauge-fixing terms. It is a usual practice in ordinary quantum field theory to impose the asymptotic boundary conditions in the path integral because it corresponds to the situation in which asymptotic states do exist and physical and nonphysical degrees of freedom in these states could be separated from each other. The asymptotic boundary conditions ensure gauge invariance of the path integral, as well as of the theory as a whole. They are an important ingredient of the BFV approach, since the Fradkin -- Vilkovisky theorem, which states the independence of the path integral on the choice of gauges, is true under the asymptotic boundary conditions.

But the asymptotic boundary conditions are not justified in the theory of gravity where the only case of a gravitating system with asymptotic states is the case of asymptotically flat spacetime. In all other cases the gravitating system does not possess asymptotic states, therefore, the asymptotic boundary conditions {\it cannot be imposed}. Taking into account this feature of the gravitational theory we shall consider the path integral {\it without} the asymptotic boundary conditions. And this is the second point where our approach differs from the one proposed by Batalin, Fradkin and Vilkovisky. In the absence of the boundary conditions the procedure of derivation of the Wheeler -- DeWitt equation is not consistent. A rigorous mathematical procedure results in the Schr\"odinger equation rather than in the Wheeler -- DeWitt one. Also, in the absence of the asymptotic boundary conditions {\it one cannot prove} gauge invariance of the theory based on the Wheeler -- DeWitt equation, and this equation loses its sense.

It is worth mentioning that some attempts to prove gauge invariance have been made. Gauge invariance means that the whole theory including a procedure of derivation of the equations for the wave function of the Universe, which are gravitational constraints in an operator form, and, as a consequence, the wave function must not depend on a choice of the lapse and shift functions, $N$ and $N^i$. In \cite{Barv} a covariant operator realization of the gravitational constraints is explored using geometrical structure of the superspace manifold. A metric on the superspace can be defined by contracting the DeWitt supermetric with some lapse function. However, the lapse function cannot be chosen arbitrary. As was shown in \cite{Barv}, only under the choice $N=1$ a desired operator realisation can be obtained. And, though the resulting equations seem to be covariant, the whole procedure of their derivation is not. In other words, if one made another choice for $N$, one would get another form of operator gravitational constraints.

An important advantage of the path integral approach is that it does not require to construct the Hamiltonian form of the theory since an equation for the wave function can be derived directly from the path integral in the Lagrangian form. The Lagrangian formalism corresponds to the original (Einstein) formulation of the gravitational theory. In what follows we shall deal with minisuperspace models to illustrate the main ideas as it is usually done in the literature (see, for example, \cite{Hall}). In these models one freezes all but a finite number of gravitational degrees of freedom. Also one supposes that spacetime has some symmetry (isotropy or, at least, homogeneity), so that one can put the shift functions $N^i=0$. We shall consider the effective action with gauge-fixing and ghost terms as it appears in the path integral approach to gauge field theories. For a model with a finite number of degrees of freedom it reads
\begin{equation}
\label{action}
S=\!\int\!dt\,\left[\displaystyle\frac12 g_{ab}(N, q)\dot q^a\dot q^b-U(N, q)
  +\pi\left(\dot N-\frac{\partial f}{\partial q^a}\dot q^a\right)
  +N\dot{\bar\theta}\dot\theta\right].
\end{equation}
Here $q=\{q^a\}$ stands for physical variables and $N$ is the lapse function, $\theta$, $\bar\theta$ are the Faddeev -- Popov ghosts. Originally, the ghosts appeared in the Faddeev -- Popov method of regularization of the path integral for gauge fields to save the unitarity property of the $S$-matrix \cite{FP}. The Batalin -- Vilkovisky effective action \cite{BV} for the Lagrangian formulation of the theory of gravity is reduced to the Faddeev -- Popov action. The gauge condition for $N$,
\begin{equation}
\label{gauge}
N=f(q)+k;\quad
k={\rm const},
\end{equation}
where $f(q)$ is an arbitrary function of physical variables, can be presented in a differential form,
\begin{equation}
\label{diff_gauge}
\dot N=\frac{\partial f}{\partial q^a}\dot q^a.
\end{equation}
The gauge condition (\ref{diff_gauge}) introduces into the effective Lagrangian the missing velocity $\dot N$, so that the Lagrange multiplier $\pi$ of the gauge condition plays the role of the momentum conjugate to $N$. Though, as was mentioned above, we do not need the Hamiltonian formulation to obtain the equation for the wave function, now we are able to construct a Hamiltonian by the usual rule
\begin{equation}
\label{Ham1}
H=p_a\dot q^a+\pi\dot N+\bar{\cal P}\dot\theta+\dot{\bar\theta}{\cal P}-L.
\end{equation}
$\bar{\cal P}$, $\cal P$ are ghost momenta. The terms with $\dot N$ are reduced in (\ref{Ham1}), so we shall come to the expression
\begin{eqnarray}
\label{Ham2}
H&=&\frac12g^{ab}p_a p_b+\pi p_a f^{,a}+\frac12\pi^2 f_{,a}f^{,a}-U(N, q)+\frac1N\bar{\cal P}{\cal P}\nonumber\\
&=&\frac12G^{\alpha\beta}P_{\alpha}P_{\beta}+U(N, q)+\frac1N\bar{\cal P}{\cal P},
\end{eqnarray}
where
\begin{equation}
\label{Galphabeta}
f_{,a}=\frac{\partial f}{\partial q^a};\quad
G^{\alpha\beta}=\left(
\begin{array}{cc}
f_{,a}f^{,a}&f^{,a}\\
f^{,a}&g^{ab}
\end{array}
\right);\quad
Q^{\alpha}=(N,\,q^a);\quad
P_{\alpha}=(\pi,\,p_a).
\end{equation}
Eq. (\ref{Ham2}) gives the Hamiltonian in {\it extended phase space}.

In his seminal paper \cite{Feyn} Feynman proposed a method of derivation of the Schr\"odinger equation from the path integral in the Lagrangian form. This method was generalized by Cheng \cite{Cheng}. Its application to cosmological models with a finite number of degrees of freedom, which are constrained systems, required further generalization that has been presented in our early work \cite{SSV1,SSV4}. The method consists in the approximation of the path integral on equations obtained by variation of the effective action (the so-called {\it skeletonization} of the path integral). The asymptotic boundary conditions aims at excluding gauge-noninvariant terms in the action and the equations, but the path integral becomes ill-definite.

In the absence of the boundary conditions the form of the Schr\"odinger equation depends on a chosen gauge condition through the function $f(q)$ in (\ref{gauge}). It reads as
\begin{equation}
\label{SE1}
i\,\frac{\partial\Psi(N,q,\theta,\bar\theta;\,t)}{\partial t}
 =H\Psi(N,\,q,\,\theta,\,\bar\theta;\,t),
\end{equation}
where
\begin{equation}
\label{H}
H=-\frac1{2M}\frac{\partial}{\partial Q^{\alpha}}MG^{\alpha\beta}
    \frac{\partial}{\partial Q^{\beta}}+U(N, q)-V[f]
   -\frac1N\frac{\partial}{\partial\theta}
    \frac{\partial}{\partial\bar\theta};
\end{equation}
the operator $H$ corresponds to the Hamiltonian in extended phase space (\ref{Ham2}), $M$ is the measure in the path integral. The feature of the procedure of derivation of the Schr\"odinger equation is the appearance of a quantum correction to the potential $U$ which has the form $\xi R$ where $R$ is the scalar curvature of configurational space with the metric $g_{ab}$, $\xi$ is some constant. For the first time the correction appeared in the work by Cheng \cite{Cheng} with $\xi=1/6$. In our case $V[f]$ denotes this quantum correction. Since the metric $g_{ab}(N, q)$ depends on the lapse function $N$, the correction will depend on the gauge-fixing function $f(q)$ after substitution of (\ref{gauge}) (see below Eq.(\ref{phys.H})).

The wave function is defined on extended configurational space with the coordinates $N,\,q,\,\theta,\,\bar\theta$, but gauge and ghost variables can be easy separated from physical degrees of freedom, since the general solution to the equation (\ref{SE1}) can be written as
\begin{equation}
\label{GS1}
\Psi(N,\,q,\,\theta,\,\bar\theta;\,t)
 =\int\Psi_k(q,\,t)\,\delta(N-f(q)-k)\,(\bar\theta+i\theta)\,dk.
\end{equation}
Here the function $\Psi_k(q,\,t)$ describes a state of the physical subsystem for a reference frame fixed by the condition (\ref{gauge}). The equation for $\Psi_k(q,\,t)$ can be obtained by substituting (\ref{GS1}) into (\ref{SE1}). The presence of the $\delta$-function in (\ref{GS1}) leads to the replacement of $Q^0=N$ by $f(q)+k$ in the operator (\ref{H}). In this way we shall get the physical Hamiltonian operator $H_{(phys)}[f]$:
\begin{equation}
\label{phys.H}
H_{(phys)}[f]=\left.\left(-\frac1{2M}\frac{\partial}{\partial q^a}M g^{ab}\frac{\partial}{\partial q^b}
 +U(N, q)-V[f]\right)\right|_{N=f(q)+k}.
\end{equation}
Separating the ghost variables we shall come to the equation for $\Psi_k(q,\,t)$:
\begin{equation}
\label{phys.SE}
i\,\frac{\partial\Psi_k(q;\,t)}{\partial t}
 =H_{(phys)}[f]\Psi_k(q;\,t).
\end{equation}
This equation can be called the Schr\"odinger equation for the physical part of the wave function. It is of particular interest for quantum cosmology.

In extended phase space approach the $0\choose 0$-Einstein equation (the Hamiltonian constraint) is replaced by $H=E$, where $H$ is the Hamiltonian (\ref{Ham2}) and $E$ is a conserved quantity. Its quantum version is the stationary Schr\"odinger equation,
\begin{equation}
\label{stat.EPS.SE}
H\Psi(N,\,q,\,\theta,\,\bar\theta)=E\Psi(N,\,q,\,\theta,\,\bar\theta).
\end{equation}
Again, substituting (\ref{GS1}) into (\ref{stat.EPS.SE}) one will come to the stationary Schr\"odinger equation for the physical part of the wave function,
\begin{equation}
\label{stat.SE}
H_{(phys)}[f]\Psi_k(q)=E\Psi_k(q).
\end{equation}
The temporal and stationary Schr\"odinger equations (\ref{phys.SE}), (\ref{stat.SE}) are a direct consequence of our method of construction of Hamiltonian dynamics in extended phase space when the lapse function $N$ becomes a dynamical variable by means of the gauge condition in a differential form (\ref{diff_gauge}).

Let me note that the role of gauge degrees of freedom in general relativity differs from their role in other gauge theories, such as electrodynamics. In these theories observables do not depend on gauge transformations, but in general relativity distances, time intervals, frequencies, and all tensor components do depend on a chosen reference frame. As well known, observed physical phenomena also depend on the reference frame. One cannot get a solution to the Einstein equations in its final form without fixing a reference frame. So, gauge invariance in general relativity is quite subtle: having a solution to the Einstein equations in some reference frame, one can recalculate it in another reference frame. To find the solution implies to determine all components of the metric tensor including gauge degrees of freedom. One cannot just neglect them or put them equal to zero because they contribute to spacetime structure.

The choice of the reference frame is equivalent to introducing $(3+1)$ spacetime splitting (see, for example, \cite{MM}). Gauge gravitational degrees of freedom can be presented by the lapse and shift functions which have a clear geometrical interpretation that was discussed by many authors. Misner, Thorne and Wheeler in their volume ``Gravitation'' \cite{MTW} described $(3+1)$ splitting as a family of spacelike hypersurfaces, each of them corresponding to some value of a time parameter $t$ (see \S 21.4). They note that to give the geometry on two successive hypersurfaces by no means fixes spacetime structure in between, and one needs also ``cross-connectors''. The lapses and shifts of the connectors determine the metric of four-geometry. Without giving the lapse and shift functions spacetime would be just a heap of hypersurfaces in no way related to each other. Again, gauge invariance of general relativity manifests itself in that one can choose another $(3+1)$ splitting and give other values to the lapse and shift functions. But one {\it has to give} some values to them. In this sense one can say that gauge degrees of freedom determine a backbone of spacetime (or ``the rigidity'' of its structure, according to Misner, Thorne and Wheeler). If one ignored them, one would obtain a theory of space but not a theory of spacetime that takes place in the Wheeler -- DeWitt quantum geometrodynamics. So, one can doubt if the Wheeler -- DeWitt theory is an attempt to quantize general relativity or some other theory of gravity in which gauge degrees of freedom play no role.

Let me remind the definition of the reference frame according to Landau and Lifshitz (\cite{LL}, \S 82):
\begin{quote}
``\ldots This result essentially changes the very concept of a system of reference in the general theory of relativity, as compared to its meaning in the special theory. In the latter we meant by a reference system a set of bodies at rest relative to one another in unchanging relative positions. Such systems of bodies do not exist in the presence of a variable gravitational field, and for the exact determination of the position of a particle in space we must, strictly speaking, have an infinite number of bodies which fill all the space like some sort of `medium'. Such a system of bodies with arbitrarily running clocks fixed on them constitutes a reference system in the general theory of relativity.''
\end{quote}
In the next paragraph Landau and Lifshitz clarify the sense of gauge invariance in general relativity:
\begin{quote}
``In connection with the arbitrariness of the choice of a reference system, the laws of nature must be written in the general theory of relativity in a form which is appropriate to any four-dimensional system of coordinates (or, as one says, in `{\it covariant}' form). This, of course, does not imply the physical equivalence of all these reference systems (like the physical equivalence of all inertial reference systems in the special theory). On the contrary, the specific appearances of physical phenomena, including the properties of the motion of bodies, become different in all systems of reference.''
\end{quote}
It is worth comparing this notion of the reference frame with the idea put forward by Einstein in his lecture ``Ether and the Theory of Relativity'' \cite{Ein}  delivered in Leiden University in 1920. Einstein emphasized that it is superfluous to postulate ether in electrodynamics, though on the other hand there is a weighty argument in favour of the ether hypothesis. He said,
\begin{quote}
``\ldots To deny the ether is ultimately to assume that empty space has no physical qualities whatever.''
\end{quote}
And later on,
\begin{quote}
``\ldots This space-time variability of the reciprocal relations of the standards of space and time, or, perhaps, the recognition of the fact that `empty space' in its physical relation is neither homogeneous nor isotropic, compelling us to describe its state by ten functions (the gravitation potentials $g_{\mu\nu}$), has, I think, finally disposed of the view that space is physically empty. But therewith the conception of the ether has again acquired an intelligible content although this content differs widely from that of the ether of the mechanical undulatory theory of light. The ether of the general theory of relativity is a medium which is itself devoid of all mechanical and kinematical qualities, but helps to determine mechanical (and electromagnetic) events.''
\end{quote}
Therefore, Einstein used the notion of ether in an absolutely new sense as some medium very closed by its qualities to that implied by Landau and Lifshitz. Recalling the concept of physical vacuum as it had been developed by theoretical physics by the end of the twentieth century, we can suppose that gravitational vacuum is a good candidate for such a medium though its properties have not been enough studied yet. The quantity $E$ in (\ref{stat.EPS.SE}), (\ref{stat.SE}) can be associated with the energy of gravitational vacuum.

However, the next question is: Can $E$ have a non-zero value? It is accepted that positive energy of matter fields balances negative gravitational energy. Zero value of $E$ is necessary to obtain the correct classical limit. But we can assume that $E$ might have had a non-zero value after the Universe creation. This possibility will be discussed in the next Section.

\section{An outline of a cosmological scenario in the extended phase space approach to quantization of gravity}
In 1980s one of widely discussed question in quantum cosmology was the question about the state ``Nothing'' from which the Universe was presumably created. How to describe this state? What wave function corresponds to it? Vilenkin \cite{Vil1} suggested that this wave function is a solution to the equation
\begin{equation}
\label{Vil_eq}
\frac{d^2\Psi}{d a^2}+\frac p a\frac{d\Psi}{d a}-\lambda^2 a^2\Psi=0,
\end{equation}
where $\lambda$ is a parameter depending on a system of units, $p$ represents the ordering ambiguity. The question about the creation of the Universe from ``Nothing'' was discussed also in our paper \cite{SSV4} where it was argued that the equation (\ref{Vil_eq}) can be thought of as the stationary Schr\"odinger equation with $E=0$ from which all degrees of freedom but the scale factor $a$ are excluded. If $p=1$ the solution we sought for is
\begin{equation}
\label{Nothing}
\Psi_0(a)=CK_0\left(\frac{\lambda a^2}2\right).
\end{equation}
This solution is the modified Bessel function of zero order. The wave function (\ref{Nothing}) describes a state with no matter and no time, keeping in mind the properties of the modified Bessel function one can see that the Universe is ``locked'' in the singularity $a=0$. These very properties enabled us to interpret this solution as corresponding to the state ``Nothing''. In \cite{SSV4} it was proposed that the creation of the Universe is a result of spontaneous reduction of this state potentially containing all the possible states of the Universe to one of them. Relative probabilities of transitions from the state ``Nothing'' to physical states are given by projections of the latter ones on the state ``Nothing''.

Although the idea of spontaneous transitions is widely used in modern physics, we should confess that it cannot give a profound explanation of the phenomenon under consideration, and we would like to have as detailed description of the phenomenon as possible. And in what state did the Universe appear right after its creation?

Some scientists assume that our Universe is not unique. Plenty of universes are created from nothing, but most of them shrink to zero size. This point of view gave rise to the approach known as ``third quantization'' (see, for example, \cite{HM}). In this approach the state ``Nothing'' is associated with the vacuum of universes, the nucleation and disappearance of a universe correspond to its creation and annihilation. Developing the analogy with particle creation, a universe may have a non-zero energy after its nucleation which it returns when disappearing, if one admits the validity of the uncertainty principle at this level. Then we come to the question, how can a tiny universe avoid shrinking and reach a macroscopic size? To answer the question we should appeal to the idea of inflation.

There are different ways to obtain the inflationary stage. Some authors just introduce the cosmological constant or model it by means of a scalar field. Another way was pointed out by Weinberg \cite{Weinberg} who shown that the Einstein equations with the cosmological constant could be derived by the variation procedure from the gravitational action under the additional condition that the determinant of the metric tensor is not a dynamical variable or, alternatively, under the condition $\det ||g^{\mu\nu}||=1$. In minisuperspace models the latter condition reduces to $N a^3=1$, where $N$ is the lapse function, $a$ is the scale factor. In fact, in this model it is the gauge condition for the lapse function. This scheme can be naturally realized in the framework of the extended phase space approach.

In \cite{Shest2} a model has been considered which is very similar to those discussed in \cite{HH,Vil1,Vil2}. This is the model of a closed isotropic universe filled with radiation. The action for this model is
\begin{equation}
\label{action-1}
S=\!\int\!dt\,\left[-\frac12\frac{a\dot a^2}N+\frac12Na
  -\frac Na\varepsilon_0
  +\pi\left(\dot N-\frac{d f}{d a}\dot a\right)
  +N\dot{\bar\theta}\dot\theta\right]
\end{equation}
Here matter fields are described phenomenologically, by the part of the action (\ref{action-1})
\begin{equation}
\label{act-mat}
S_{(mat)}=-\!\int\!dt\frac Na\varepsilon_0.
\end{equation}
The Schr\"odinger equation for the physical part of the wave function reads
\begin{equation}
\label{Schr-eq-gen}
\left.\left[-\frac12\sqrt{\frac N a}\frac d{d a}
  \left(\sqrt{\frac N a}\frac{d\psi}{d a}\right)
 +\frac12 N a\psi-\frac Na\varepsilon_0\psi+V[f]\psi\right]\right|_{N=f(a)}=E\psi.
\end{equation}

The model considered in \cite{HH} includes a conformally invariant spatially homogeneous scalar field which, after redefinition
$\phi\rightarrow\displaystyle\frac{\phi}a$ and rescaling, gives the following contribution to the action:
\begin{equation}
\label{act-scal}
S_{(scal)}=\frac12\int\!dt\left(\frac aN\dot\phi^2-\phi^2\right).
\end{equation}
The model with the scaler field leads to the Schr\"odinger equation
\begin{equation}
\label{Schr-eq-scal}
\left.\left[-\frac12\sqrt{\frac N a}\frac{\partial}{\partial a}
  \left(\sqrt{\frac N a}\frac{\partial\Psi}{\partial a}\right)
 +\frac12 N a\Psi
 +\frac12\frac Na\left(\frac{\partial^2\Psi}{\partial\phi^2}-\phi^2\Psi\right)
 +V[f]\Psi\right]\right|_{N=f(a)}=E\Psi.
\end{equation}
Variables in this equation can be easily separating by demanding that
\begin{equation}
\label{Psi-sep}
\Psi(a,\,\phi)=\psi(a)\chi(\phi),
\end{equation}
where $\chi(\phi)$ satisfies the equation
\begin{equation}
\label{chi-eq}
-\frac12\frac{\partial^2\chi}{\partial\phi^2}+\frac12\phi^2\chi=\varepsilon_0\chi,
\end{equation}
so that for $\psi(a)$ we would come to Eq.(\ref{Schr-eq-gen}). The matter in (\ref{action-1}), (\ref{Schr-eq-gen}) is represented by a medium with the equation of state $p_{(mat)}=\displaystyle\frac13\varepsilon_{(mat)}$.

At the same time, the gauge-fixing term in the action can be interpreted as describing another medium (another subsystem) whose state is determined by a chosen gauge condition. One can say that in \cite{HH} the condition $N=a$ was implicitly chosen. However, we are interested in the condition $N a^3=1$. In this case the equation of state of this subsystem is $p_{(gauge)}=-\varepsilon_{(gauge)}$.

The form of the Schr\"odinger equation, as well as the effective potential, is determined by the chosen gauge condition. Here we shall present a qualitative analysis without quantitative estimation. In particular, we shall ignore the term $V[f]$ since this quantum correction has a little effect on the potential. We traditionally consider here models of a closed universe, as it was done in the first works on quantum cosmology, though nothing prevent us from analysing open models. The main goal of this analysis is to demonstrate that the extended phase space approach admits evolutionary development of the wave function, in contrast to the Wheeler -- DeWitt quantum geometrodynamics in which the wave function describes the only state of the Universe unchanging with time.

\begin{figure}
\centering
\includegraphics{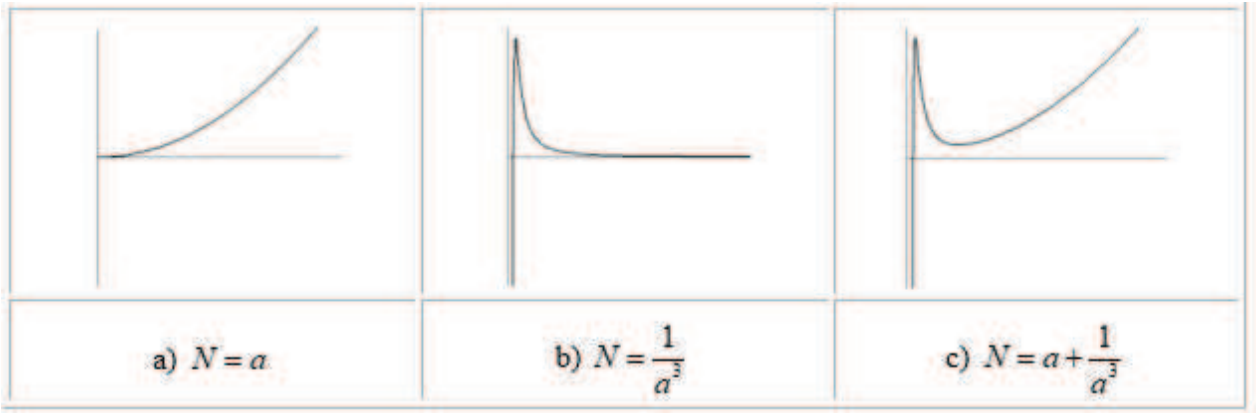}
\caption{\protect\small The effective potentials \label{f1}}
\end{figure}

The forms of the potential for various gauge conditions are given at Fig.~\ref{f1}.

Fig. 1(a) corresponds to a Friedmann closed universe in which expansion can be replaced by contraction. The case depicted at Fig. 1(b) is an inflationary universe. One can suppose that the Universe was created in a metastable state in the region of very small values of the scale factor $a$ and then tunneled through the potential barrier. The effective potential $U(a)=\displaystyle\frac1{2a^2}-\frac{\varepsilon_0}{a^4}$ depends on the parameter $\varepsilon_0$. The smaller the parameter is, the higher and narrower the barrier becomes. There is a non-zero probability for the Universe to reach arbitrary large values of the scale factor, it means that the Universe may expand to infinity. The probability of tunnelling can be calculated in accordance with quantum mechanical rules.

The case depicted at Fig. 3(c) is more realistic. It covers the two previous cases as asymptotic limits. Again, the Universe might be created in the region of very small values of the scale factor and then tunneled to the region of large enough values of $a$. As the scale factor increases, the inflationary stage would be replaced by the slow Friedmann expansion.

If we limit ourself by consideration only stationary states, we would hardly achieve better results than those in the Wheeler -- DeWitt quantum geometrodynamics since the Universe in a stationary state with some fixed value of $E$ does not evolve. To describe processes in the Early Universe we should admit some factors which make the Universe change its state.

If the Universe were born in the region of small $a$, there would exist two possibilities. The first one is that the Universe would shrink to a point returning the non-zero energy $E$ to the state ``Nothing''. The second possibility is that the Universe would tunnel through the barrier or go over it if the value of $E$ is bigger than the height of the barrier. As was mentioned, the height of the barrier is determined by the parameter $\varepsilon_0$ which characterizes the energy of matter fields. In this model we presume the existence of matter fields in the Early Universe, therefore, we should take into account particle creation in the expanding Universe \cite{GP}. Particle creation acts as a perturbation altering the effective potential and stimulating quantum transitions between the levels of $E$. One can expect that as a result of quantum transitions the Universe will be in the state with the minimal value of $E$ that can be set to zero.

In the case of $E=0$ the Schr\"odinger equation for the physical part of the wave function formally coincides with the Wheeler -- DeWitt equation. However, its form will depend on the chosen gauge condition, so, this equation cannot express gauge invariance of the theory. Let us remind the well-known fact that the Wheeler -- DeWitt equation is not invariant with respect of the choice of parametrization of gauge variable. In \cite{SSV2} it has been demonstrated that the parametrization noninvariance of the Wheeler -- DeWitt equation is equivalent to its gauge noninvariance.

With these remarks in mind, one can claim that the Wheeler -- DeWitt equation is valid at the stage when particle creation can be neglected. Of course, in principle, the Universe can be nucleated in the state with $E=0$, but we do not have enough grounds to discard non-zero values of $E$. As has been already emphasized, in the absence of asymptotic states one cannot ensure gauge invariance of quantum gravity. It is reasonable to expect that gauge-noninvariant effects manifest themselves stronger in the Very Early Universe and weaken as the Universe expands. From the point of view of the extended phase space approach the Wheeler -- DeWitt equation is legitimate in the period immediately preceding the stage when the Universe is good enough described by classical general relativity. The Wheeler -- DeWitt equation has a correct quasiclassical limit. So, one can see that in this approach the Universe passes through a series of states in the course of its evolution. Starting from the state in which the Universe is localized in the region of small $a$ with probably non-zero value of $E$, as a result of quantum transitions the Universe turns out to be in the state that is described by a solution to the Wheeler -- DeWitt equation, and in the present stage its state is described by a quasiclassical wave function.

The model we have considered is very simplified. More sophisticated models can include various matter fields and extra dimensions. In the above the existing of matter in the Very Early Universe has been assumed, that is a necessary condition for particle creation and a trigger for quantum cosmological evolution. However, one can put the question about the origin of matter in the Universe right after its nucleation. Let us mention the idea of multidimensional gravity that gravitational degrees of freedom corresponding to extra dimensions may be interpreted in some cases as gauge or matter fields, then gravity might be a source of all physical fields and interactions.

\section{On the choice of gauge conditions}
In the above consideration the choice of the gauge condition plays an important role. The choice of the condition $Na^3=1$ is equivalent to introducing a cosmological constant into the Einstein equations, so it assures the inflationary stage in the course of the Universe evolution. Thus, in the approach presented here the choice of gauge conditions implies the choice of a model.

Alternatively, one can construct a model with a scalar field with required properties to get the inflationary stage. These two ways lead to the same result at least from a formal mathematical point of view. The first variant seems to be paradoxical since one believes that the choice of a model is a search for an objective description of nature, but the theory must by no means depend on the choice of gauge conditions.

The gauge condition determines the dependence of time scale on geometric characteristics of space, in a certain sense, it determines the spacetime structure. Introducing the gauge condition we make an assumption on spacetime structure in the Early Universe. Similarly, the choice of the model with the scalar field means that at the classical level we have selected solutions to the Einstein equations with suitable qualities. At the quantum level the gauge condition affects the form of the Schr\"odinger equation for the physical part of the wave function, in its turn, it specifies possible quantum states of the Universe in the beginning of its evolution and, eventually, its present state. So it seems that by the choice of the gauge condition we affect the past of the Universe.

Let me proposed an analogy between this situation and the Wheeler delayed-choice experiment \cite{Wheel}. As it is known, in the Wheeler thought experiment a photon emitted millions years ago encountered a very massive galaxy acting like a gravitational lens. When the photon arrived at the Earth, an astronomer can decide what he would observe: the photon passed either to the left or to the right of the galaxy or an interference picture. And again it seems that the choice of the type of observation determines which way the photon followed in the past or if it was in a superposed state.

To my mind, the Wheeler experiment can be understood in the best way if one appeals to the path integral formalism. In this formalism one should integrate over all possible ways of the photon. However, if one wish to take into account what type of the experiment was chosen, it may give some limitations on admissible ways of the photon to integrate over.

In quantum gravity one should introduce some gauge conditions in the path integral. The gauge conditions also imply some limitations on possible ways of the Universe evolution, on admissible cosmological scenarios. Our speculations have not been confirmed by real observations. But by observing the Universe today we deal with some kind of delayed-choice experiments. These experiments demonstrate that properties of a quantum object depend on its whole history. In attempting to explain the present Universe we can make assumptions about conditions in its past. If we consider the whole Universe as a quantum object we should not be surprised that its properties in the past are influenced by its present.
It may look paradoxical but it is in accordance with the spirit of quantum theory.

\section{Conclusions}
The central feature of the extended phase space approach is the possibility to describe the evolution of the Universe including its quantum stage as developing in time. At this point the proposed approach is radically differs from the Wheeler -- DeWitt quantum geometrodynamics where the Universe is supposed to be in the only state described by a time-independent wave function. Moreover, some authors advocate the idea that time is something irrelevant in physics (see, for example, \cite{RV}).

In the extended phase space approach the meaning of the wave function is close to that in quantum mechanics. It describes different states of the Universe which correspond to different stages of its evolution. Of course, time is a classical notion. But, as Bohr emphasized, we should describe phenomena in a convenient language. Abandoning the notion of time we can hardly speak about relations among quantum phenomena in the Early Universe and find out how they can determine initial conditions for classical evolution. We believe that a logically consistent and profound quantum theory of gravity will be constructed. But we definitely need time to achieve it.

\section*{Acknowledgments}
I am grateful to the organizers of the IV International Conference on Particle Physics and Astrophysics for the opportunity to give a talk.

\end{document}